\documentclass[useAMS,usenatbib]{mn2e}
\usepackage{epsfig}

\def\lesssim{\mathrel{\hbox{\rlap{\hbox{\lower4pt\hbox{$\sim$}}}\hbox{$<$}}}}
\def\gtrsim{\mathrel{\hbox{\rlap{\hbox{\lower4pt\hbox{$\sim$}}}\hbox{$>$}}}}
\def\la{\mathrel{\hbox{\rlap{\hbox{\lower4pt\hbox{$\sim$}}}\hbox{$<$}}}}
\def\ga{\mathrel{\hbox{\rlap{\hbox{\lower4pt\hbox{$\sim$}}}\hbox{$>$}}}}

\def\spose#1{\hbox to 0pt{#1\hss}}
\def\approxlt{\mathrel{\spose{\lower 3pt\hbox{$\sim$}}
	\raise 2.0pt\hbox{$<$}}}
\def\approxgt{\mathrel{\spose{\lower 3pt\hbox{$\sim$}}
	\raise 2.0pt\hbox{$>$}}}

\def\<{\thinspace}

\def\boxit#1{\vbox{\hrule\hbox{\vrule\kern3pt\vbox{\kern3pt
          #1 \kern3pt}\kern3pt\vrule}\hrule}}

\def\deg{^o}

\def\ga{{\rm\thinspace gauss}}

\def\h50{\hbox{$\rm\thinspace h_{50}$}}
\def\h50m1{\hbox{$\rm\thinspace h_{50}^{-1}$}}

\title[Exploring the Properties of the M31 Halo Globular Cluster System]
{Exploring the Properties of the M31 Halo Globular Cluster System}
\author[Huxor et al.]{A. P. Huxor$^{1}$\thanks{Present address: H.H.Wills Physics Laboratory, Tyndall Avenue, Bristol, BS8 1TL}, A. M. N. Ferguson$^{1}$,  N. R. Tanvir$^{2}$, M. J. Irwin$^{3}$,  A. D. Mackey$^{1}$\thanks{Present address: Research School of Astronomy \& Astrophysics, Australian National University, Mt. Stromlo Observatory, Cotter Road, Weston Creek, ACT 2611, Australia},
\newauthor  R. A. Ibata$^{4}$, T. Bridges$^{5}$, S. C. Chapman$^{3}$, G. F. Lewis$^{6}$\\
$^{1}$Institute for Astronomy, University of Edinburgh, Royal Observatory, Blackford Hill, Edinburgh EH9 3HJ\\
$^{2}$Department of Physics and Astronomy, University of Leicester, University Road, Leicester LE1 7RH\\
$^{3}$Institute of Astronomy, Madingley Road, Cambridge, CB3 0HA \\
$^{4}$Observatoire de Strasbourg, 11, rue de l'Universite, F-67000, Strasbourg, France \\
$^{5}$Department of Physics, Queen's University, Kingston, Ontario, Canada K7M 3N6 \\
$^{6}$Sydney Institute for Astronomy, School of Physics, A29, University of Sydney, NSW 2006, Australia \\
}

\begin{document}

\date{}

\pagerange{\pageref{firstpage}--\pageref{lastpage}} \pubyear{2007}

\maketitle

\label{firstpage}

\begin{abstract}
  Following on from our discovery of a significant population of M31
  outer halo globular clusters (GCs), and updates to the Revised
  Bologna Catalogue of M31 GCs, we investigate the GC system of M31
  out to an unprecedented radius ($\approx 120$~kpc). We derive
  various ensemble properties, including the magnitude, colour and
  metallicity distributions, as well as the GC number density profile.
  One of our most significant findings is evidence for a flattening in
  the radial GC number density profile in the outer halo.
  Intriguingly, this occurs at a galactocentric radius of $\sim$ 2
  degrees ($\sim$ 30 kpc) which is the radius at which the underlying
  stellar halo surface density has also been shown to flatten.  The
  GCs which lie beyond this radius are remarkably uniform in terms of
  their blue (V$-$I)$_0$ colours, consistent with them belonging to an
  ancient population with little to no metallicity gradient.
  Structural parameters are also derived for a sample of 13
  newly-discovered extended clusters (ECs) and we find the lowest
  luminosity ECs have magnitudes and sizes similar to Palomar-type GCs
  in the Milky Way halo.  We argue that our findings provide strong
  support for a scenario in which a significant fraction of the outer
  halo GC population of M31 has been accreted.
\end{abstract}


\section{Introduction}

The properties of globular cluster (GC) systems provide valuable
probes of the formation and evolution of their host galaxies (e.g.
\citealt{Westetal04,BrodieStrader06}).  It is commonly believed that
GCs form in major star-forming episodes that accompany galaxy
formation, as well as in subsequent merger events (e.g.
\citealt{zepfash93}).  Furthermore, the native GC population of a
galaxy is expected to be augmented through mergers with and accretions
of smaller systems, each of which will bring its own retinue of GCs
into the final galaxy.  As a result, the GC population of a galaxy
will reflect both the amount of mass formed in situ, as well as that
which has been accreted.

As GCs are (mostly) luminous and compact, they can be readily observed
in galaxies up to a few hundred Mpc distant. Nevertheless, the GC
systems of Local Group galaxies remain of central importance as they
allow the most detailed study of the properties of GC populations and
how they correlate with the formation history of their host galaxies.
This is possible as their proximity permits studies of resolved field
and cluster stellar populations through their colour-magnitude
diagrams (e.g.  \citealt{Mackeyetal06}) and spectroscopy
(e.g.\citealt{Barmbyetal00}).

The study of the Galactic GC system by \citet{SearleZinn78} proved
crucial for understanding the history of our own Milky Way (MW).
Their analysis of GC metallicities within the halo led them to
conclude that many of these objects must have formed within
protogalactic fragments that fell into the Galaxy after the collapse
of the central regions had been completed.  This scenario was in stark
contrast to the monolithic slow collapse picture earlier proposed by
\citet{els62}.  One of the main distinguishing characteristics of the
two models was the halo metallicity gradient; the latter scenario
predicted a radial gradient with clusters nearer to the centre being
more metal-rich, as they were formed somewhat later in the collapse,
while the former predicted a spread in metallicities at all radii but
no significant radial variation. Modern theories of galaxy assembly
are based on hierarchical structure formation with galaxies forming
inside dark matter halos (e.g.  \citealt{wf91}).  Although this model
bears some similarities to the scenario proposed by Searle \& Zinn,
there are also several differences.  In particular, it has been shown
that build-up via accretion can sometimes lead to halo metallicity
gradients since massive satellites, which are normally more
metal-rich, sink further into the potential well of the host than do
low mass objects (e.g.  \citealt{font08,delucia08}).  Thus, while the
lack of a halo metallicity gradient argues against slow
pressure-supported collapse, the existence of one could be consistent
with both that scenario as well as accretion.

The GC system of M31 has also been studied intensively in order to
search for clues about how that system formed and evolved (e.g.
\citealt{Cramptonetal85,Elson88,Huchraetal91,
  Barmbyetal00,Perrettetal02,Fanetal08}). Since M31 is similar to the
MW in many respects, it may be expected to have experienced a similar
assembly history. Early work suggested a mild gradient in the
metallicity of the M31 GC system (e.g. \citealt{Sharov88,
  Huchraetal91,Barmbyetal00}.  \citet{Perrettetal02} and
\citet{Fanetal08} found that this result is primarily driven by a
metallicity gradient in the metal-poor GCs alone, while
\citet{Perrettetal02} noted that the slope of the gradient may flatten
beyond a projected radius of $\sim$ 14 kpc.

In addition to the metallicity gradient, another key property of a GC
system is the radial profile - the areal number density of GCs as a
function of distance from the centre of the host galaxy.  Radial GC
profiles have traditionally been represented with either a power-law
(i.e.  $R^{-n}$) or a de Vaucouleurs law (i.e. $R^{1/4}$ law), with a
flattening of the profile at small radii, and a gradual steepening in
the outer regions \citep{BrodieStrader06}.  The most recently
published GC surface density profile in M31 is that of
\citet{Battistinietal93}, who experimented with a variety of fitting
formulae. Using a sample which extended to a radius of $\sim30$~kpc,
they noted that their data were consistent with a steepening in the
outer regions, earlier found by \citet{Racine91}, if one uses the
general power -- or R$^{1/4}$ laws (although this is not the case if
they employ an R$^{1/1.6}$ law).  They argued that this steepening was
unlikely to be due to incompleteness in their sample at large radius,
although the possibility could not be completely excluded.

A major limitation of all previous studies of the ensemble properties
of the M31 GC system has been the restricted radial range of the
samples.  These studies have generally employed samples extending to
no more than $\sim25$~kpc, while analogous studies of the MW GC system
have extended to beyond 100 kpc.  In the last decade, vast amounts of
new imaging have been obtained of the outer regions of M31 revealing
that both field stars and GCs extend to radii of well over 100 kpc
(e.g.  \citealt{Fergusonetal02,Ibataetal07, Huxoretal08,
  McConnachieetal09,Mackeyetal10a}).  It is therefore timely to
revisit the properties of the M31 GC system in light of these new
datasets. In Section 2, we discuss the sample used for our study which
consists of the revised Bologna catalogue and new GCs presented in
\citet{Huxoretal08} (hereafter Paper I).  In Section 3, we discuss
radially-dependent properties of the M31 GC system using a baseline
that extends over 100~kpc. In Section 4, we discuss properties of
extended star clusters in M31's outer halo.  Section 5 discusses our
findings in the context of the assembly history of M31.
 
\section{The sample}

\begin{figure}
 \centering
 \includegraphics[angle=0,width=90mm]{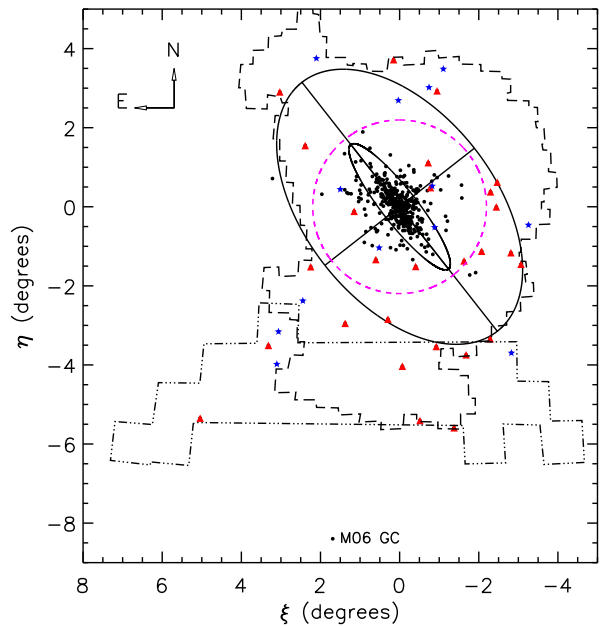} 
 \vspace{2pt}
 \caption {The location of the new globular clusters (red triangles)
   and extended clusters (blue stars) in relation to confirmed RBC GCs
   that lie in the survey area analysed here (black circles).  The
   limited extent of the known GC population prior to this study can
   be clearly seen.  The dashed line outlines the INT survey area
   covered, and the dashed-dotted line outlines that part of the
   Megacam survey employed here.  The inner ellipse has a semimajor
   axis of 2$^\circ$ (27 kpc) representing a disk with an inclination
   of 77.5$^\circ$; the optical disk of M31 lies well within this
   boundary.  The outer ellipse denotes a flattened ellipsoid of
   semi-major axis length 4$^\circ$ (55 kpc).  The dashed circle lies
   at a radius of 30 kpc, and shows the region at which the break in
   the GC surface density profile occurs (see text for details).The
   outermost GC found by \citet{Martinetal06} and discussed by
   \citet{Mackeyetal10a} is also shown (M06 GC) and indicates the
   extent of the currently-known M31 halo GC
   population.}\label{Fi:plot} \end{figure}

The sample of GCs used in this study is taken from version
V3.5\footnote{http://www.bo.astro.it/M31/} of the Revised Bologna
Catalogue (RBC), which was released in March 2008\footnote{While this
  paper was in preparation, the RBC was updated in December 2009, to
  V4.0.  However the changes do not affect the results in the present
  work, as the updates concern a few GCs in the inner region of M31
  and/or add new young clusters. We do not include such young clusters
  from our sample (see main text).}.  A full description of the RBC
can be found in \citet{Galletietal04} with updates in
\citet{Galletietal06} and \citet{Galletietal07}. It contains 1983
entries, including 509 confirmed GCs, 13 confirmed extended clusters
(ECs) and 1049 ``candidate clusters".  This version of the RBC
  includes 103 confirmed objects reported by \citet{Kimetal07}
  although subsequent work has shown that many of these objects have
  been improperly classified \citep{Caldwelletal09,Peacocketal10}.
  This has only a minor impact on the results presented here as the
  Kim et al.  objects all lie within a projected radius of 18~kpc
  (90\% within 10~kpc) from the centre of M31, while our main focus in
  this paper is on the halo regions beyond 20~kpc. The RBC V3.5 also
includes the 40 new outer halo clusters found by our group and
reported in Paper I.  In brief, these latter objects were discovered
in more than 80 square degrees of imaging survey data taken with the
Wide-Field Camera on the Isaac Newton Telescope and Megacam on the
Canada France Hawaii Telescope (see \citealt{Fergusonetal02}, and
\citealt{Ibataetal07}, for details of the surveys).  Thanks to the
quality of the seeing conditions and the proximity of M31 to us, the
clusters could all be identified as marginally or fully resolved
stellar concentrations.  The new clusters we discovered increased the
total number of confirmed GCs in M31 known at the time of publication
by a modest amount ($\sim$ 10\%) but the number of confirmed GCs known
beyond 1$\degr$ ($\approx 14$~kpc) by more than 75\%.

For the main sample used in our analysis, we select clusters in the
RBC V3.5 that have a \emph{Global classification flag} (f) equal to 1
or 8 corresponding to 'confirmed' GCs and ECs respectively. Although
the RBC uses this flag to distinguish between ECs and GCs, studies to
date suggest M31 ECs have the same ancient stellar content as GCs
\citep{Mackeyetal06}, hence we consider them together in the present
work. The very remote GC found by \citet{Martinetal06} and recently
studied in detail by \citet{Mackeyetal10a} is included in the sample
for most of the analysis but not for the GC surface density profile as
it lies in an area of the CFHT/Megacam survey for which full results
of our GC search have yet to be published.  Possible \emph{young} GCs
are also excised from our sample by removing those clusters for which
the \emph{young cluster} flag (yy) is greater than zero.  While M31
appears to have a genuine population of young to intermediate age GCs
(e.g. \citealt{FusiPeccietal05,Fanetal06}), such a population is
absent in the MW. Since our primary interest in this paper is to
examine ancient star clusters as probes of galaxy formation, a sample
with these young objects removed makes for a more appropriate
comparison of the GC systems of both galaxies.  The final sample we
employ for the bulk of our analysis contains 431 objects while the
sample used for constructing the surface density profile analysis
contains 430 objects - removing the \citet{Martinetal06} GC as noted
above.

The RBC is a compilation of a variety of different data sources and in
situations where our own photometry is available from
\citet{Huxoretal08}, this is used in preference to that provided by
the RBC for homogeneity. The projected galactocentric radii of the GCs
are re-derived from the equatorial coordinates listed in the RBC using
M31 central coordinates of RA$=00^h42^m44.3^s$ and Dec$=+41\deg 16'
09^{\prime\prime} $ for the centre of M31, taken from
NED\footnote{http://nedwww.ipac.caltech.edu/}.  Figure \ref{Fi:plot}
shows the spatial distribution of the GCs around M31 in our sample and
illustrates the much greater radial extent of our sample compared to
those used for previous M31 GC studies.

 \section{Radially-Dependent Properties of the M31 GC System} 

 We now proceed to use this sample to derive the basic properties of
 the M31 GC system out to much larger radius than has been previously
 done. Wherever possible, we compare the quantities derived with those
 of the MW GC system for which we use the information listed in the
 McMaster Catalogue of Milky Way Globular Clusters
 \footnote{http://physwww.physics.mcmaster.ca/$\sim$harris/mwgc.dat}
 \citep{Harris96}. This catalogue lists data for 141 GCs, although not
 all of them have a full set of derived parameters. The catalogue also
 does not contain a few very recently-discovered MW GCs in the
 Galactic plane (e.g. \citealt{Kobulnickyetal05, Kurtevetal08}) but
 since these do not lie at large galactocentric distances, they do not
 affect our overall results. We do, however, include the two
 newly-discovered GCs found in the outer halo of the MW
 \citep{Koposovetal07}.  As there is limited information available for
 these objects at the present time (only size and V-band magnitudes),
 they can only be used for a subset of the following comparisons.  The
 values of E(B-V) given in the Harris catalogue are used to derive the
 corrected magnitudes and colours for each MW cluster.

\subsection{GC Luminosities} 

\begin{figure} 
 \includegraphics[angle=90,width=90mm]{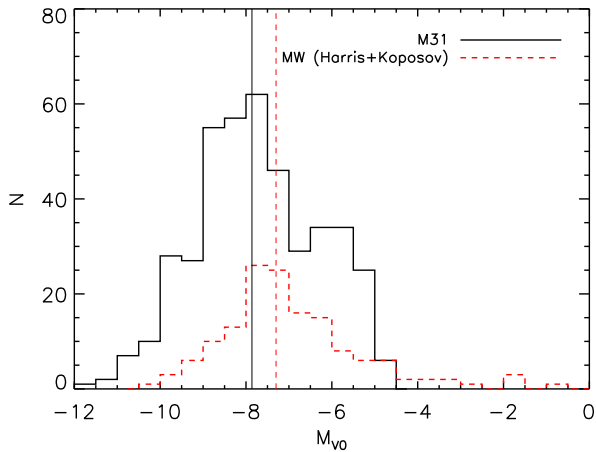}
 \vspace{1pt}
 \caption{Extinction-corrected absolute V-band magnitudes of M31 GCs
   (black line) in our catalogue, using values of A$_{V}$ derived
     from \citet{Fanetal10} or \citet{Schlegeletal98}, depending on
     availability (see text for details). The red dashed histogram
     shows the distribution for the MW GCs, and includes the
     \citet{Koposovetal07} clusters. Median values for each sample are
     shown with vertical lines (-7.9 for M31 and -7.3 for the MW).
   \label{Fi:Mv_histo}}
\end{figure}

The M31 GC luminosity function (GCLF) is shown in Figure
\ref{Fi:Mv_histo}.  The M31 GC magnitudes have been calculated
assuming a distance modulus of 24.47 mags \citep{McConnachieetal05}.
The magnitudes are corrected for extinction using the E(B-V)
  values in \citet{Fanetal10} where available.  These values are
  derived from SED fits to multi-band photometry and give the total
  line-of-sight extinction including that internal to M31. For those
  GCs in our sample that were not studied by \citet{Fanetal10}, we
  adopt their median value of E(B-V) = 0.12, except for the very outer
  halo GCs (R$_{proj} > $40 kpc), where we use only Galactic
  foreground extinction calculated by interpolation of the
  \citet{Schlegeletal98}
  maps\footnote{http://astro.berkeley.edu/$\sim$marc/dust/data/data.html}.
  Previous work on the M31 outer halo GCs \citep{Mackeyetal07} gives
  reddening values consistent with Galactic extinction being the
  significant component. The MW GCs are overlaid in red.

Although the number of currently-confirmed M31 GCs is already $\sim 3$
times larger than the number known in the MW, the shapes of their
luminosity functions are roughly similar, at least down to an absolute
magnitude of M$_{V0}\approx -5$, where completeness issues begin to
complicate the M31 sample.  Figure \ref{Fi:Mv_histo} reveals an offset
of $\sim$ 0.6 mags between the peaks of the GCLFs (the medians being
-7.9 and -7.3 for M31 and the MW respectively) however the magnitude
of this offset is very dependent on the assumed extinction for the M31
population.  While the \citet{Fanetal10} reddening values are the best
available as they are derived in a homogeneous manner for a large
number of clusters, there are hints that they may overestimate the
extinction at high E(B-V) when compared to other methods (see their
Figure 7) .  Indeed, if only Galactic foreground extinction is
assumed, there is no discernible offset in the GCLF peaks of M31 and
the MW. The M31 GCLF also exhibits a secondary peak at
  M$_{V0}\approx -5.5$ that is not seen in the MW population.  This
  peak is present when inner ($\le 20$~kpc) and outer ($\ge 20$~kpc)
  halo samples are considered separately, suggesting it could be
  genuine. On the other hand, many of the \cite{Kimetal07} clusters
  have magnitudes around this value, a large fraction of which have
  questionable classifications (e.g. see the discussion in
  \citet{Peacocketal10}).  A more detailed comparison of the MW and
  M31 GCLFs will be carried out at a later date, when uniform
  photometry and reddenings are available for the entire sample and
  the completeness and contamination of the M31 catalogue has been
  more rigorously quantified.

\begin{figure} 
 \includegraphics[angle=90,width=90mm]{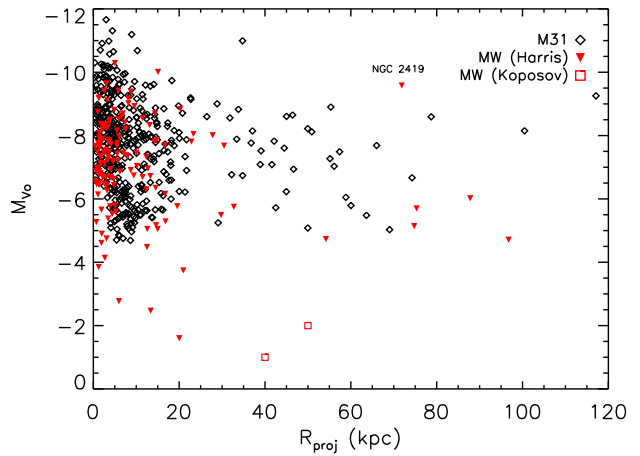}
 \vspace{1pt}
 \caption{The radial variation of the extinction-corrected V-band
   absolute magnitudes for M31 (open diamonds) and MW (inverted
   triangles and open squares) GCs, using the same data as Fig.
   \ref{Fi:Mv_histo}. In the case of the MW GCs the actual distance
   (R$_{gc}$) is converted to an ``average projected distance" through
   R$_{proj}$ = R$_{gc}\times (\pi/4)$. The unusual MW cluster NGC
   2419 is labeled.  \label{Fi:bothMvR}}
\end{figure}

Figure \ref{Fi:bothMvR} shows the distribution of absolute magnitudes
against projected galactocentric distance for those GCs for which
V-band data is available. This includes all but four of the Harris MW
catalogue and all but eight of the M31 catalogue.  The galactocentric
distances for the MW GCs, $R_{gc}$, are converted to equivalent
projected radii using $R_{proj} = \pi/4 \times R_{gc}$.  This
numerical factor accounts for the ratio of the projected distance to
the true distance for an isotropically-distributed GC population
viewed from a distant random external vantage point.

Figure \ref{Fi:bothMvR} reveals that M31 has a population of luminous
GCs at large galactocentric distances, a finding previously commented
on from studies with smaller datasets
\citep{Mackeyetal07,Galletietal07}. While these GCs are no more
luminous than GCs at smaller radii in M31, they are significantly more
luminous than the outer halo GCs in the MW. In the MW, only NGC 2419
has a galactocentric radius larger than $\sim$ 35~kpc, and a
luminosity greater than M$_{V0} \lesssim$ -6.5 whereas M31 has 21 GCs
in this same region of parameter space (even although the sample
studied here covers only a quarter of the sky area at these
distances). Almost all the outer halo GCs in the MW are
``Palomar-type" clusters with low luminosities and relatively large
half-light radii. As both samples are likely to be highly complete in
this magnitude range, the difference between the numbers of luminous
M31 and MW GCs, a factor of $\sim$ 20, cannot be solely due to the
difference in the overall size of the GC populations (which accounts
for only a a factor of $\sim$3).  It is unclear whether M31 also
possesses an excess of faint outer halo GCs compared to the MW since
completeness issues currently complicate our analysis below M$_{V0}
\approx$ -5.

\subsection{GC Colours and Metallicities}

\begin{figure} 
 \includegraphics[angle=90,width=90mm]{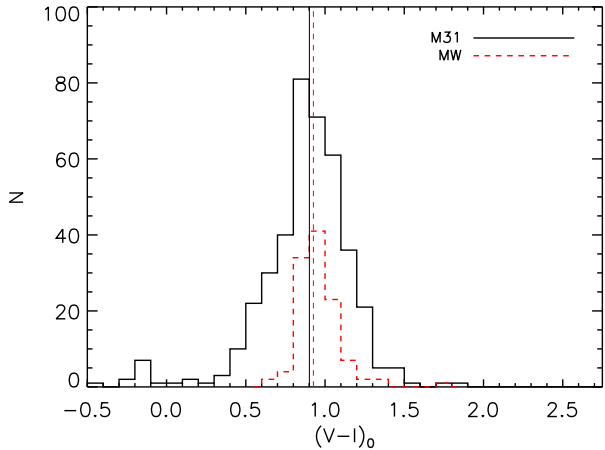}
 \vspace{2pt}
 \caption{The distribution of (V-I)$_{0}$ for GCs in M31 (black solid
   histogram) and the MW (red dashed histogram), where this colour
   measure is available. Where no I band data are available for MW
   GCs, (V-I) colours are derived from (B-V) (see text for details).
   Median values are shown by vertical lines. The
   \citet{Koposovetal07} GCs are not included here as no colour
   information is available for them. \label{Fi:histogram_vi}}
\end{figure}

Fig \ref{Fi:histogram_vi} shows histograms of the (V-I)$_0$ colours
for the MW and M31 GC samples.  Only 96 out of the 141 MW GCs had
(V-I) values listed in the Harris database.  As an additional 20 MW
GCs had (B-V) data, we fit the (V-I) colours as a function of (B-V)
for those GCs for which both colours were available. This resulted in
the relation (V-I) = 1.23 $\times$ (B-V) + 0.09 (RMS = 0.05), which
was used to estimate (V-I) values for those remaining 20 GCs. 
  Figure \ref{Fi:histogram_vi} shows that the M31 GC system has the
  same median colour as that of the MW, with the reddening-corrected
  median (V-I)$_0$ colours being 0.89 and 0.93 mags for M31 and the MW
  respectively. This is somewhat at odds with earlier findings that
  the M31 GC system is, on average, redder than that of the MW (e.g.
  \citealt{Huchraetal91}) and depends largely on our adopted reddening
  values from \citet{Fanetal10}.

\begin{figure} 
 \includegraphics[angle=90,width=90mm]{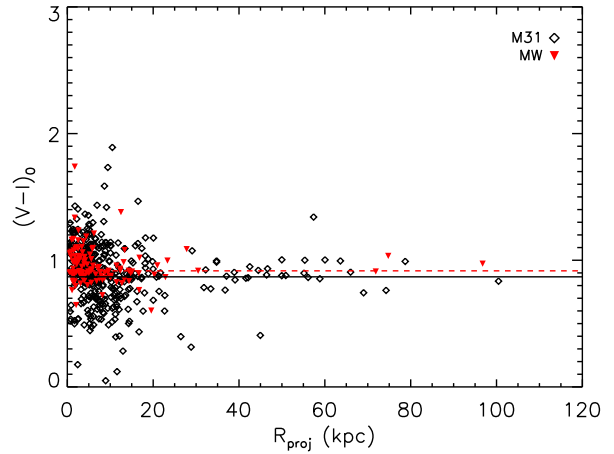}
 \vspace{2pt}
 \caption{The distribution of (V-I)$_{0}$ for M31(open diamonds) and
   MW (inverted triangles) with galactocentric radius.  The data used
   here is the same as for Fig \ref{Fi:histogram_vi}. The mean colours
   for the R$_{proj}$ $> $30 kpc M31 and R$_{proj}$ $ >~15$ kpc MW GCs are
   shown as black and red dashed lines
   respectively.}\label{Fi:plot_r_vi}
\end{figure}

When viewed as a function of galactocentric radius (Figure
\ref{Fi:plot_r_vi}), the mean GC colours in the outer halo are
uniformly blue and similar in both systems.We find ((V-I)$_0 =
  0.87\pm 0.04$ in M31 for R$\gtrsim 30$~kpc and (V-I)$_0 = 0.92\pm
  0.03$ for R$\gtrsim 15$~kpc in the MW, where errors are the standard
  error on the mean. The colours are almost identical for R$\gtrsim
45$~kpc, the difference for R$\gtrsim 30$~kpc solely due to the
inclusion of the very blue extended cluster HEC1. As small-scale
extinction variations are not likely to be a factor in these remote
parts, this suggests that intrinsic properties of the outer halo GCs
in both systems are very similar.  Interestingly, the mean colour of
the outer halo GCs in M33 ((V-I)$_0 = 0.88\pm0.05$
\citealt{Huxoretal09} is also very close to the values found in M31
and the MW.

\begin{figure} 
  \includegraphics[angle=90,width=90mm]{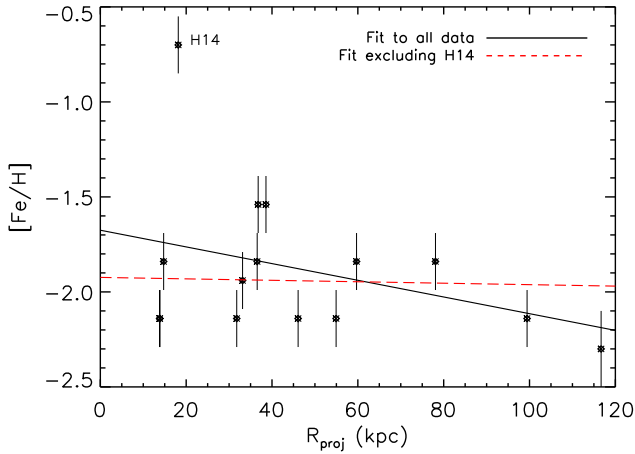}
  \vspace{2pt}
  \caption{The radial variation of metallicities for those GCs with
    accurate values in the literature from CMD fitting
    \citet{Mackeyetal06,Mackeyetal07,Mackeyetal10a}. The solid line is
    a fit to all the data, and the dashed line is a fit that excludes
    the outlier, the possibly intermediate-age cluster H14
   ( \citet{Mackeyetal07}, where is it called GC7). The errors on individual data points are
    $\pm$0.15 dex
    \citep{Mackeyetal06,Mackeyetal07,Mackeyetal10a}.\label{Fi:dougal_1}}
\end{figure}

The dispersion in colour of the outer halo GCs in Figure
\ref{Fi:plot_r_vi} is relatively small beyond 30~kpc and there is no
evidence of any radial gradient. Since the integrated colour reflects
both the age and the metallicity of a GC, further information is
required in order to properly interpret the uniformity in the outer
halo.  For several of the M31 outer halo GCs, we have previously
derived metallicities from fitting model isochrones and MW globular
cluster fiducials to high-quality colour-magnitude diagrams
\citep{Mackeyetal06,Mackeyetal07,Mackeyetal10a}. These fits are based
exclusively on the red giant and horizontal branch stars.  Figure
\ref{Fi:dougal_1} shows the metallicities derived in this manner for
11 compact and 4 extended GCs in the M31 halo.  A linear least-squares
fit to all the data yields a slope of -0.006 $\pm$ 0.002 dex
kpc$^{-1}$, indicating a marginal negative gradient. However there is
one clear outlier in the plot -- H14 -- which was previously flagged
by \citet{Mackeyetal07} as a possible intermediate-age cluster. If
true, this would affect the metallicity of the cluster which has been
derived assuming a 10~Gyr age. If this object is removed, we obtain
the dashed line fit with a mean metallicity of -1.94 $\pm$ 0.22 dex
and no discernible gradient (-0.003 $\pm$ 0.002 dex kpc$^{-1}$).  If
we focus only on those GCs which lie beyond 30~kpc, the mean
metallicity is effectively the same, given the errors, at -1.95 dex.
It thus seems fair to assume that the uniformity in the broadband
colours of the outer halo GCs seen in Fig.  \ref{Fi:plot_r_vi}
reflects a narrow spread in both age and metallicity, with this
population being predominantly old and metal-poor.  For comparison,
the average [Fe/H] for the MW sample with R$_{proj} >~15$ kpc is
-1.70$\pm0.23$ dex \citep{Harris96}. Our finding of no
  significant metallicity gradient in the M31 halo GC population is
  consistent with the study of \citet{AlvesBritoetal09}, who found the
  same result from spectroscopic measurements of a small sample of
  GCs, although we note that there is a discrepancy between the
  absolute metallicities obtained by these authors and those inferred
  from our CMD fitting).

\subsection{The Radial Surface Density Profile of the M31 GC System}

The radial number density profile of the M31 GC system can be
calculated in a straightforward fashion once a correction is made for
the non-uniform azimuthal coverage of the outer halo surveys from
which our GC catalogue has been constructed. Indeed, as can be seen in
Figure \ref{Fi:plot}, the surveys used to identify outer halo GCs
extend significantly further in the south eastern direction than
anywhere else.  Using knowledge of the field centres and sizes, we
determined the proportion of the sky imaged within a specific annulus
and hence the correction factor needed to estimate the number of
clusters within that radial range (on the assumption of an isotropic
distribution).  Our number densities are calculated within circular
annuli, which should be appropriate if the true halo shape is roughly
spherical.

\begin{figure} 
 \centering
 \includegraphics[angle=90,width=90mm]{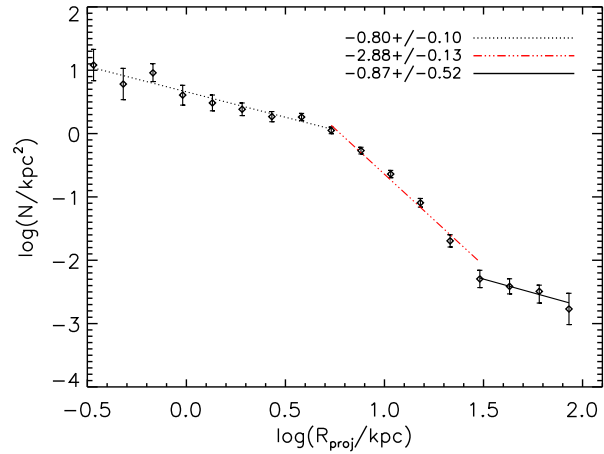}
 \vspace{2pt}
 \caption{A log-log plot of the radial number density profile of GCs
   in M31, with Poisson errors (see text). A broken power-law has been
   fit, with slopes and errors for each component given in the legend.
   The inner region extends from the centre of M31 to a projected
   distance (R$_{proj}$) of $\approx$ 5 kpc (dotted black line); the
   intermediate region from R$_{proj}$ $\approx$ 5 to $\approx$ 30 kpc
   (red dot-dashed line); and the outer region from R$_{proj}$
   $\approx$ 30 kpc (solid black line).  A change in the slope of the
   profile can be seen at $\approx$ 5 and 30 kpc. The outer three bins
   contain 13, 10 and 3 GCs in the original data which increase to
   estimated values of $\approx$ 15, 25 and 27 GCs when the incomplete
   spatial coverage for each annulus is taken into account.}
 \label{Fi:profile}
\end{figure}

\begin{figure} 
 \centering
 \includegraphics[angle=90,width=90mm]{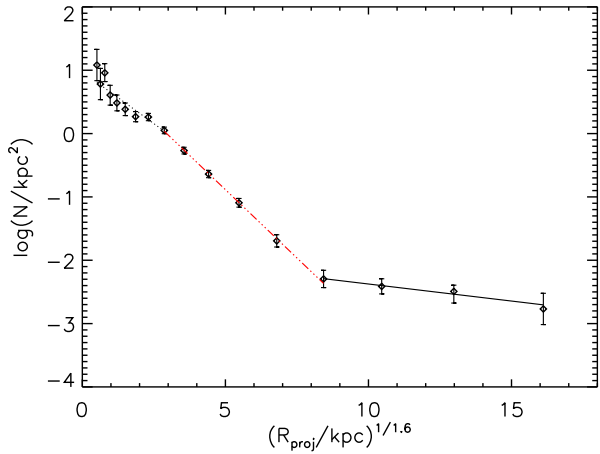}
 \vspace{2pt}
 \caption{The radial profile of surface density of GCs in M31 against
   R$^{1/1.6}$, using the same bins as for Figure \ref{Fi:profile}.
   Linear fits have again been made to the inner, intermediate and
   outer regions as in Fig. \ref{Fi:profile}.  The flattening in the
   last three bins remains very clear. }
 \label{Fi:profile_batt}
\end{figure}

Figure \ref{Fi:profile} shows the projected GC number density as a
function of radius for M31, with the correction for incomplete
azimuthal coverage included.  The data are presented in equally-spaced
logarithmic bins. The $\pm1\sigma$ errors are determined assuming the
number of GCs at a given radius is a Poisson process. Then in each
bin, the upper (lower) end of the error bar is set to the value of
underlying GC density for which the probability of observing a number
of clusters as low (high) as we actually did, or lower (higher), is
15.87\% \citep{Mulder83}.

The profile shows a broken power-law behaviour. Inside a projected
radius of $\sim 5$~kpc, the radial number density profile of GCs is
rather flat, as has been previously commented on in earlier studies
(e.g.  \citealt{deVauc78,Harris79,Wirth85, Battistinietal93}). While
this behaviour could be genuine, it could also represent
incompleteness in the samples at small radii.  Indeed, detecting GCs
against the high surface brightness bulge which dominates in this
region remains a challenge even for modern day surveys.  Beyond this
inner region, the profile is much steeper out to a projected radius of
$\sim 30$~kpc after which it flattens again.  This is the first time
that this behaviour has been seen in M31 and is due to the fact
that the new profile extends to radii of $\sim$ 100 kpc, more than
three times further out than the earlier profiles of \cite{Racine91}
and \cite{Battistinietal93}. 

A broken power-law fit is overlaid in Figure \ref{Fi:profile} with
best-fit indices of $-0.80 \pm 0.10$ inside of $\sim 5$~kpc, $-2.88
\pm 0.13$ in the range $5\lesssim \rm{R_{proj}} \lesssim 30$~kpc and
$-0.87 \pm 0.52$ outside of $\sim 30$~kpc.  Although this power-law
representation is typically employed in extragalactic GC studies, it
is clearly a rather poor description of the behaviour in M31,
especially at intermediate radii, being too steep at small radii and
too shallow at large radii in this range.  \citet{Battistinietal93}
also noted that power-law fits were a poor representation of the
radial number density profile of their sample of M31 GCs and
investigated a number of other empirical fits.  They found most
success with an R$^{1/\it{m}}$ law, with $\it{m}$ $ \sim 1.6 $.  In
Fig. \ref{Fi:profile_batt}, we show a profile of this form fit to our
current sample.  Even though there have been many updates to the M31
GC catalogue since Battistini's work, this form still provides an
excellent description of the radial number density profile over a
significant radial range ($5 \lesssim \rm{R_{proj}} \lesssim 30$~kpc).
Indeed, this relation even provides a good fit for the outer halo GC
profile, albeit with a different slope.

It is worth commenting that while the outer halo region is
  particularly rich in the more extended clusters (described below),
  these objects are not the primary cause of the flattening in the
  profile.  Of the 36 GCs beyond the break in the GC surface density
  profile, only nine are ECs. This can be compared to 2 ECs from the
  36 GCs in an annulus immediately interior to the break.  If the
  radial surface density profile is constructed using only compact
  GCs, the the break is still clearly present, but the slope at large
  radii is somewhat steeper.
 
\section{The Extended Cluster Population}

One of the main results thus far from our GC search has been the
discovery of a population of ECs within the halo of M31 (Paper I and
\cite{Huxoretal05}). These objects typically have half-light radii of
$\gtrsim20$ pc and are significantly more extended than the normal GC
population which have half-light radii of a few pc.
\cite{Huxoretal05} and \cite{Mackeyetal06} presented an analysis of
the four brightest ECs discovered in the early stages of our survey.
Here we present structural parameters for all thirteen ECs in the
sample studied here.

To investigate the structures of these clusters, we consider empirical
King profiles,

 \[
   \Sigma (r) = \Sigma _{0} \left[  \frac{1}{\left(1 + \left( r/R_{c} 
	\right)^2 \right)^{1/2}} - \frac{1}{\left(1 + \left( R_{t}/R_{c} 
	\right)^2 \right)^{1/2}}       \right]^2         
 \]  
 
 \noindent where $R_{c}$ and $R_{t}$ are the core and tidal radii
 respectively.  These profiles were fit to V band (in the case of INT)
 or g- band (for Megacam) photometry which were the shortest
 wavelength data available to us.  \citet{King85} notes that the best
 cluster profiles are obtained by using the bluest images since
 statistical fluctuations due to individual red giants can be
 problematic at longer wavelengths. The fits were made to the
 cumulative flux measured within a series of apertures of increasing
 radii made with the IRAF/apphot photometry package\footnote{IRAF is
   distributed by the National Optical Astronomy Observatories, which
   are operated by the Association of Universities for Research in
   Astronomy, Inc., under cooperative agreement with the National
   Science Foundation.}.  Hence the fit, which minimised $\chi^{2}$,
 was made to the integral of the King profile, which gives the total
 flux within a radius r:
 
\[
C(r) = \Sigma _{0} 2 \pi \left[ \frac{R_{c}^{2}}{2}\ln\left(\alpha \right)
- \frac{2R_{c}^{2}}{\beta}\left(\alpha \right)^{1/2} 
+ \frac{r^{2}}{2\beta^{2}} + \frac{2R_{c}^{2}}{\beta}
	\right]        
\] 

\noindent  where 

\[
\alpha = 1 + \left( r/R_{c} \right)^2 	    
\] 

\noindent and 

\[
\beta = \left(1 + \left( R_{t}/R_{c} \right)^2 \right)^{1/2}   	    
\] 
 
There are some uncertainties associated with the values derived from
our King profile fits.  In the case of HEC7, contamination from
foreground stars was removed first by using the ``patch" option in
GAIA\footnote{GAIA, the `` Graphical Astronomy and Image Analysis"
  tool is now available as part of the Starlink Software Collection,
  via http://starlink.jach.hawaii.edu/} to replace the affected region
with an average of the background sky. Another cluster, HEC4, lies at
the edge of a CCD chip but the fractional area missing is small
allowing us to still obtain photometry to a radius of 13 arcsec.
Cluster HEC6 was so heavily affected by background galaxies that no
profile fit could be attempted.  In this case, a value for $R_{h}$ was
obtained by finding the aperture containing half the luminosity as
given by our measurement of the total magnitude within a 12 arcsec
aperture, and thus has a significant degree of uncertainty.

\begin{table*}
 \centering
 \caption[Properties of the Extended Clusters]{
   Properties of the Extended Clusters. Columns 2 -- 5 and 8 -- 9 derived by fits to a King model.
   Columns 6 and 7 show values of $R_{c}$ and $R_{h}$ derived from
   recent HST/ACS imaging of HEC4,5,7 and 12, indicating errors in the ground-based data.  The values for HEC6 are not derived from model-fitting, see text for details. The cross identifications for
   the ECs in \citet{Huxoretal05} and Tanvir et al. (submitted)  are also given (H05 ID and T10 ID, respectively) where appropriate.
 }
  \begin{tabular}{@{}llllllllll@{}}
     HEC ID & $R_{c}$  & $R_{t}$  & $R_{h}$  & $M_{V0}$ & H05 ID & T10 ID& $R_{c} $HST  & $R_{h} $HST   & $M_{V0} $HST  \\
           &  (pc) &  (pc) & (pc) & (model)  & &  &   (pc) &  (pc) &   (pc) \\
 \hline 
  1 & 18 & 113 & 24   & -6.0   & - & -& - & -  & -  \\
  2 & 12 & 83 & 17 & -5.3 & - & - & - & -&  - \\
  3 & 13 & 116 & 21 & -4.5 & - & - & - & -   &  -  \\
  4 & 16 & 140   & 26   & -7.2 & C3 & EC3  & 25.9 & 18.3 & -7.45 \\
  5 & 23  & 166   & 34  &  -7.3 & C1 & EC1  & 14.8 & 24.1 & -7.68 \\
  6 & -  & -   & 24  &  -5.5 & - & -  &- & - & - \\
  7 & 17  & 132  & 26  &  -7.8 & C2 & EC2 & 10.9 & 20.0 & -7.03 \\
  8 & 11 & 58 & 22  & -5.1& -  & - & - & - & - \\
  9 & 20 & 94 & 24 &  -6.0 & - & -  & - & - & -  \\
  10 & 10 & 135 & 19 & -6.3 & - & -  & - & -  & -   \\
  11 & 14 & 94 & 20 &  -6.7 & - & -  & - & -   & -   \\
  12 & 32 & 84 & 27 &  -5.4 & - & EC4  & 28.9 & 27.7 & -6.68  \\
  13 & 23 & 114 & 27 & -4.4 & - & -  & - & -  & -   \\
\hline
\end{tabular}
\end{table*}

For four of the ECs studied, we can compare the ground-based profile
fits made here with those derived from higher spatial resolution
HST/ACS data and reported in Tanvir et al. (2011). The HST data
allow for better removal of background galaxy contamination and give
profile fits that are, for three of the four ECs, somewhat smaller
than those found with the ground-based data.  It is likely that this
is mostly due to the limitations inherent in constructing profiles
from ground-based data.  This underscores the need to treat the
quantities derived for faint ECs from ground-based fits with some
caution.

\begin{figure} 
 \includegraphics[angle=90,width=90mm]{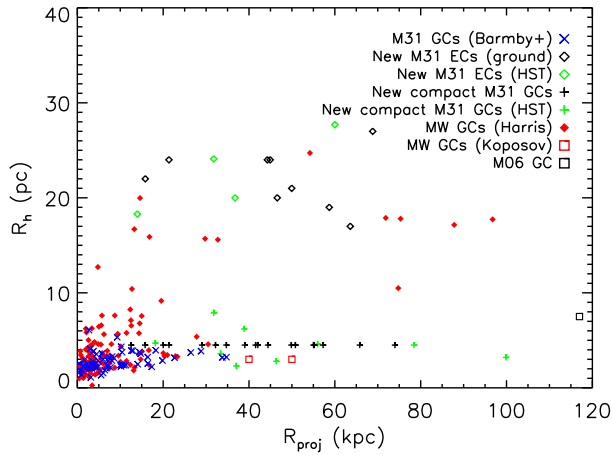}
 \vspace{2pt}
 \caption{The distribution of R$_{h}$ for M31 and MW GCs with
   projected galactocentric distance (kpc) showing the MW GCs (filled
   diamonds), new extended clusters from Paper I (open diamonds), the
   \citet{Barmbyetal07} clusters that are in our catalogue
   (triangles). The new compact GCs from Paper I (plus signs) are shown
   with an R$_{h}$ value of 4.5 pc (except for ten clusters, where values are available from Tanvir et al.
   (submitted)). This is intended to show a maximum
   value, as they are unresolved in our data. See text for details.
 }\label{Fi:Rp_Rh_plot}
\end{figure}

It is of interest to compare the size distribution of ECs with that of
the more compact GCs in M31.  Although we would have liked to obtain
profile fits for our entire sample of outer halo GCs, this proved to
be impossible for the compact objects.  The average seeing of our
ground-based data was 1.2~arcsec, which corresponds to a physical size
of $\sim$4-5 pc at the distance of M31. Thus, we do not resolve
classical GCs, which typically have core radii of 2-3 pc, to a
sufficient extent to accurately fit their radial profiles.  Instead,
we use as a comparison sample the work by \citet{Barmbyetal07} and
Tanvir et al. (2011) which presents King profile fits for compact M31
GCs from high resolution imagery with HST.  We also include the high
resolution ground-based imaging study of the outermost M31 halo GC by
\citet{Mackeyetal10a}.  Figure \ref{Fi:Rp_Rh_plot} shows the
half-light radii (R$_{h}$) distribution of M31 ECs (black (green) open
diamonds show ground-based (HST) measurements) and GCs (crosses)
against projected galactocentric radius compared to the MW GCs (filled
diamonds). We only show those members of the \citet{Barmbyetal07}
sample that are also found in our master M31 GC catalogue described
previously. Most of those in \citet{Barmbyetal07} that are not in our
list are putative ``young" clusters, as indicated by the corresponding
flag in the RBC database.  A further handful are not yet included in
the RBC.  Figure \ref{Fi:Rp_Rh_plot} includes the newly found compact
M31 GCs from Paper I (plus signs) which are plotted with an upper
limit on the half-light radius of 4.5 pc. (black crosses), derived
from the average seeing. The newly discovered GCs from Paper I with
structural parameters derived from our HST data (Tanvir et al.  2011)
are shown as green plus symbols).  It can be seen that the M31 GCs
exhibit a very different size distribution compared to those of the
MW, in that the MW does not possess compact GCs at large
galactocentric radii. There is also a suggestion of bimodality in the
size distribution of M31 GCs at large radius with few GCs having
R$_{h}$ in the range from 8 to 15 pc.

\begin{figure} 
 \includegraphics[angle=0,width=90mm]{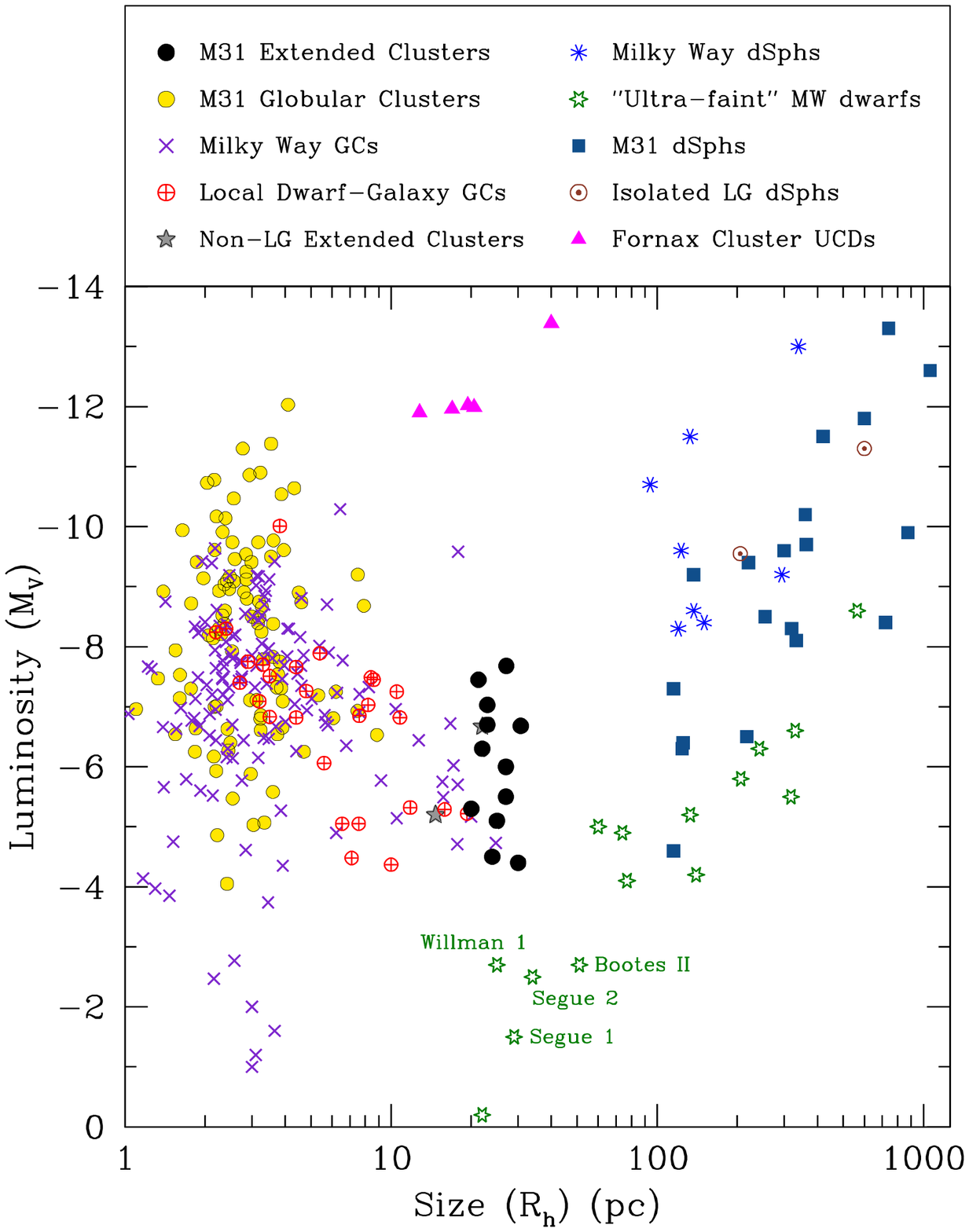}
 \vspace{2pt}
 \caption{Plot of M$_{V}$ against R$_{h}$ of the ECs (filled circles) --  with the data from Tanvir et al. (submitted) used where available,
   shown with a range of low-mass steller systems in the Local Group,
   including: M31 GCs - circles \citep{Barmbyetal07}, MW GCs - crosses \citep{Harris96,Koposovetal07}, GCs in
   LMC, SMC, Fornax \& Sag. dwarf - crossed circles
   \citep{vandenBerghMackey04, Harris96}, the ECs found beyond the Local Group - filled stars \citep{Mouhcineetal10, DaCostaetal09}, UCDs - triangles
   \citep{Mieskeetal02, DeProprisetal05, Drinkwateretal03}, M31 dwarf
   galaxies - squares
   \citep{McConnachieetal08,Zuckeretal04,Zuckeretal07,Harbecketal05,
     Irwinetal08, Martinetal06,Ibataetal07,Majewskietal07,Martinetal10}, classical
   MW dwarf galaxies - asterisks
   \citep{IrwinHatzidimitriou95,McConnachieetal06}, the islolated
   Local Group dwarfs Cetus and Tucana - dotted circles
   \citep{McConnachieetal06,Savianeetal96}, and the newly found
   low-luminosity MW dwarf galaxies - stars \citep{Belokurovetal08, Belokurovetal10},
   for which the data of \citet{Martinetal08} and \citet{deJongetal10} are used. Note that Segue 3 from \citet{Belokurovetal10} has the structural parameters of a GC, and so is plotted as a MW GC. }\label{Fi:mv_rh} \end{figure}

The ECs are of particular interest since, as first noted by
\citet{Huxoretal05}, they lie between classical GCs and dwarf
spheroidal galaxies in a plot of M$_{V}$ vs R$_{h}$ (Figure
\ref{Fi:mv_rh}).  The true nature of ECs -- whether simply star
clusters or dark-matter dominated systems -- remains unclear at
present.  \citet{Collinsetal09} present a measurement of the internal
velocity dispersion in HEC12 (EC4) which was used to derive a
mass-to-light ratio of M/L = 6.7$^{+15} _{-6.7}$
M$_{\odot}$/L$_{\odot}$.  Although consistent with a globular star
cluster, this result has large uncertainties and cannot be used to
definitively exclude the presence of a modest amount of dark matter.
The first three extended clusters reported in \citealt{Huxoretal05}
were extreme in their large magnitudes and half-light radii. While
these objects appeared to be somewhat isolated in the M$_{V}$-R$_{h}$
plot, the additional clusters reported here indicate that ECs actually
form a continuous, nearly vertical, sequence which overlaps at the
faint end with the smaller, less-luminous Palomar-type GCs found in
the MW. Moreover, the new HST data suggests that the most extreme ECs
are not as large as previously thought.  However, they are still
significantly more extended than typical GCs, and one might speculate
whether a few of M31 ECs are higher luminosity analogues of some of
the unusual ultra-faint systems that have recently been discovered
around the Milky Way, such as Segue I \citep{Belokurovetal07}.  Some
of the ultra-faints which most resemble ECs in terms of their
luminosities and sizes are labelled in Figure \ref{Fi:mv_rh}.  There
is still much debate about the true nature of the lowest luminosity
ultra-faint systems, with opinions split between their (once?) being
genuine dwarf galaxies as opposed to simple star clusters (e.g.
\citealt{Siegeletal08,NOetal09}).  In-depth imaging and spectroscopic
studies of the M31 ECs as well as the ultra-faint galaxy population
will provide further insight into this question.

\section{Discussion} 

We have presented a detailed study of the properties of the M31 GC
system, with a particular focus on the halo region, using the largest
sample compiled to date of confirmed outer halo GCs. Wherever
possible, we have compared the properties of the M31 halo GCs to those
of their counterparts in the MW, often finding marked differences.  In
particular, M31 has far more GCs than the MW (at least by a factor of
3) and hosts a population of luminous compact GCs at large radius
(R$_{proj} \lesssim 30$~kpc) that, aside from NGC~2419, is completely
absent in the Milky Way.  M31's halo GC system is also considerably
larger in physical extent than the MW's, with the most remote member
currently-known lying at a projected (3D) radius of 120~kpc
(200~kpc)\citep{Mackeyetal10a}.  It is unlikely that the difference in
the outer halo GC populations is due to intrinsic differences in the
shape of the GC luminosity functions as we have shown that these are
in good agreement, although the peak values in the two systems are
slightly offset.  The outer halo of the MW is inhabited primarily by
Palomar-type GCs which are characterized by low luminosities and
diffuse structures.  Such clusters may also exist in the outer halo of
M31 but, with typical M$_{V0} \gtrsim -5$, they are difficult to
detect at that distance.  A detailed assessment of the completeness of
our M31 GC catalogue is required before we can determine whether M31
also possesses an excess of Palomar-type GCs at large radius and
conduct a full analysis of the GCLF.

M31 hosts a population of luminous extended GCs which currently have
no known counterparts in the MW.  With M$_V \lesssim -6$ and R$_h
\gtrsim 20$pc, such objects should have been easily detected with the
SDSS out to a radius of $\sim 300$~kpc within the Galactic halo (see
Figure 10 in \citealt{Koposovetal08}).  It may be that similar objects
exist in the MW halo yet are so few in number that they lie in parts
of the sky not yet surveyed to sufficient depth.  Alternatively, if
ECs represent the tail of the size distribution of clusters at a given
magnitude, then the absence of luminous examples in the MW may simply
be due to the overall difference in the number of luminous GCs in the
two systems.  It is possible that some of the Palomar-type GCs
in the MW are the low luminosity analogues of the ECs seen in M31.  It
is of interest to note that many of the Palomar-type GCs have been
shown to have younger ages than the bulk of the halo GCs and evidence
supports the notion that many of these objects formed in dwarf
galaxies that subsequently merged with the MW (e.g.
\citealt{MackeyvandenBergh05}). On the other hand, although EC4 does
not appear to possess a dark matter halo, it cannot yet be 
ruled out that some ECs in M31 may share more of a kinship with dwarf
galaxies -- perhaps representing the bright tail of the population of
ultra-faint dwarfs that has recently been uncovered in the MW
\citep{Belokurovetal07}. Some of the ECs are very faint and
  extended (e.g. HEC13, see Table 1), approaching Willman 1 in its
  properties.

The M31 GC population has often been considered to be somewhat redder
than that of the MW but with the reddenings we have adopted here the
two systems have very similar colours. The halo GC populations
(R$_{proj} \gtrsim 15-30$~kpc) in the two systems are especially
noteworthy as they have almost identical (V-I)$_0$ colours which
remain constant with radius.  Using a subsample of M31 GCs with known
[Fe/H] from CMD-fitting, the mean metallicity of the outer halo GCs is
[Fe/H]$\sim -1.9\pm 0.2$~dex with no discernible gradient.  This is in
reasonable agreement with the mean for the outer MW GC population,
[Fe/H]$\sim -1.7\pm 0.2$~dex beyond 15~kpc, but considerably lower
than the mean of the M31 stellar halo over the same radial range,
[Fe/H] $\sim -0.8$ to $-1.4$~dex
\citep{Kaliraietal06,Chapmanetal06,Kochetal08,Richardsonetal09}. This
argues against the outer M31 field halo and halo GC system forming in
situ at the same epoch.  Furthermore, following the argument made by
\citet{SearleZinn78} in the context of the MW, the lack of a
metallicity gradient within the M31 halo GC population argues against
the system having formed as part of a pressure-supported slow
collapse.  Instead, it is consistent with the picture wherein the
outer GC system formed in a number of smaller subsystems which later
merged to form the halo.

\begin{figure} 
  \centering
  \includegraphics[angle=90,width=85mm]{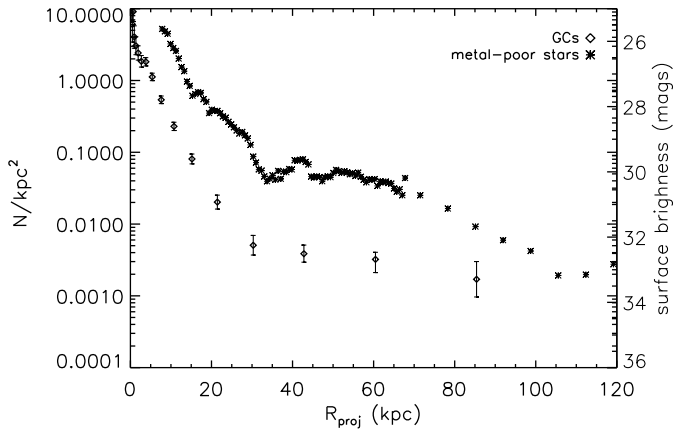}
  \vspace{2pt} \caption{Plot of the M31 GC radial profile (diamonds)
    and the arbitrarily-scaled minor-axis metal-poor ($-3.0
    <$ [Fe/H] $< -0.7$) profile
    (asterisks) from \citet{Ibataetal07}. Errors for the stellar profile are less
    than, or comparable to, the symbol size. The GC data are presented
    in the same bins as in Fig.  \ref{Fi:profile}. The flattening
    toward the central regions of M31 seen in that figure was more
    apparent than here as it employed a logarithmic axis for
    galactocentric radius.}
  \label{Fi:rod_profile}
\end{figure}

The large radial baseline spanned by our sample of GCs has enabled
construction of the GC number density profile to hitherto unprobed
distances. An unexpected result is the flattening of the profile
beyond a radius of 30~kpc (see Figures \ref{Fi:profile} and
\ref{Fi:profile_batt}). Intriguingly, a similar flattening has been
observed in the surface brightness profile of the underlying stellar
halo. This was first pointed out by \citet{Irwinetal05} who used RGB
star count data in a large swath centered on the southern minor axis
of M31 and later confirmed and extended by \citet{Ibataetal07} who
constructed an azimuthally-averaged stellar surface brightness profile
for the entire south-east quadrant of the galaxy, reaching distances
of $\sim 150$~kpc. Figure \ref{Fi:rod_profile} compares the M31 GC
number density profile with the metal-poor minor axis profile of the
stellar halo from \citet{Ibataetal07}.  Both profiles flatten at
roughly the same radius, $\sim 25-30$~kpc, however the GC number
density profile appears flatter beyond this break than that of the
field stars.  \citet{Ibataetal07} showed that, beyond a projected
radius of 30~kpc, the stellar halo profile can be fit with a power-law
of form $\Sigma_V \propto R^{-1.91\pm0.12}$ while we have found here
that the GCs behave as $\propto R^{-0.87\pm0.52}$.  It is unclear at
present whether this difference should be viewed as significant.  The
slope of the stellar surface density profile is highly dependent on
accurate subtraction of contaminating foreground stars, while the GC
number density profile suffers from small number statistics at large
radius.

The fact that both GC and field star populations reveal profile
flattenings at around the same radius is consistent with the idea that
accretion has played an important role in the formation of the outer
halo.  \citet{Abadietal06} have used numerical simulations of galaxy
formation within a $\Lambda$CDM cosmological framework to investigate
the structure of galaxies formed through a combination of in situ star
formation and accretion. They show that beyond the luminous edge of
the galaxy, defined as that radius at which the accreted stellar
component starts to dominate over that of the in-situ component, the
slope of the radial surface brightness profile changes (see their
Figure 4).  Within this radius, which corresponds to 20~kpc in their
simulations, the surface brightness profile is well fit by a de
Vaucouleurs bulge plus an exponential disk. At larger radii, the outer
halo profile flattens significantly and can be fit with a power-law
which varies from $\Sigma_V \propto R^{-2.3}$ at ~30 kpc to $\propto
R^{-2.9}$ at 100~kpc. This behaviour is qualitatively consistent with
both the outer field star and GC number density profiles in M31,
however the observed radial fall-offs are shallower than the
simulations predict (see also \citet{Ibataetal07}).  It is intriguing
that the GC number density profile in M31 has an outer slope that is
reasonably close to that expected for the dark matter halo -- $\propto
R^{-1.5}$ -- calculated over the range 25--100~kpc, using the
parameters given in \citealt{Klypinetal02}).

The lack of a metallicity gradient in the halo GC population and the
shape of the GC areal number density profile both provide support for
accretion playing a significant role in building up the M31 GC system.
The sheer number of outer halo GCs and the existence of particularly
extended clusters could also be signatures of this mode of formation.
In this scenario, the overall differences in the halo GC populations
of the MW and M31 could be the result of the two galaxies having
experienced a different number of accretion events, or accretions of a
different type. For example, the MW may have accreted mostly low mass
satellites which carry few, if any, associated GCs while M31 may have
undergone at least one more substantial merger (e.g.
\citealt{Fardaletal08}).

While ample evidence exists for satellite accretion events
contributing field stars to stellar halos
(\citealt{Ibataetal94,Ibataetal01,Fergusonetal02, McConnachieetal09}),
direct evidence for GC systems being built up in this manner has been
less forthcoming.  One notable example is the Sagittarius dwarf that
is currently being accreted onto the Milky Way and which has been
shown to be contributing at least one massive compact GC (M54) as well
as several Palomar-type clusters \citep{Bellazzinietal03,
  DaCostaArmandroff95, ForbesBridges10}.  More recently, direct
evidence for GC accretion in M31 has been presented by
\citet{Mackeyetal10b}.  These authors examine the spatial correlation
between the positions of a sample of halo GCs (many of which are
included in the present sample) and underlying tidal debris streams.
They use a Monte Carlo approach to show that the probability of the
observed degree of alignment being due to chance is low, below 1\%,
and conclude that the observed spatial coincidence reflects a genuine
physical association.  They further argue that the accretion of
cluster-bearing satellite galaxies could plausibly account for
$\gtrsim 80\%$ of the M31 GC population beyond 30~kpc.  The properties
of the M31 halo GC population presented in this paper are wholly
consistent with this idea.

Finally, it is interesting to compare the halo GC system of M31, the
most populous and extended GC system of a disk galaxy in the local
Universe, with that of the well-studied giant elliptical M87.
Wide-field ground-based studies of M87's GC population have recently
been carried out by \cite{aTamuraetal06, bTamuraetal06} and
\citet{Harris09} enabling the system to be traced out to distances of
$\gtrsim 100$~kpc.  Since GCs at the distance of M87 ($\sim 16$~Mpc)
are unresolved from the ground, these studies need to rely on
statistical subtraction to detect and characterize the GC population
which can lead to some uncertainties, particularly at large radii.
Both \citet{Harris09} and \citet{bTamuraetal06} find that the red GCs
in M87 are concentrated in the central regions of the galaxy
($\lesssim 50$~kpc) while the blue population can be traced out to at
least 100~kpc and shows no evidence for a colour gradient in these
parts. This is very similar to our finding of a large colour spread in
the inner regions of M31 with a very uniform blue population
dominating in the outer halo. The M87 studies also derive radial
number density profiles for the GCs and show a single profile form
(either R$^{1/4}$ or power law $R^{-n}$ where n smoothly increases
with radius) fits the data well out to $\gtrsim 150$~kpc. However the
profile shape at large radius is highly dependent on the assumed level
of contaminating sources (e.g. foreground stars, compact background
galaxies) and it is not yet possible to place rigorous constraints on
the form of M87's GC distribution in these parts (although the GC
  system of M87 has the benefit of not suffering from problems
  associated with small number statistics). Indeed, inspection of
Figure 5 from \citet{bTamuraetal06} hints at a possible flattening in
the number density profile at $\sim 90$~kpc, similar to what we see at
$\sim 30$~kpc in M31.  \citet{bTamuraetal06} also note that the blue
GCs in M87 are more extended than the stellar light, but recent work
by \citet{Janowieckietal10,Williamsetal07} shows M87's stellar halo is
far larger than previously thought. Using very deep wide-field
imagery, they trace the M87 surface brightness profile to $\sim
180$~kpc at which point the V-band surface brightness falls below
their detection limit of 29 magnitudes per square arcsec. This is
entirely consistent with the extent of the blue GC population. In
summary, although M87 and M31 differ vastly in terms of their inner
morphologies and galaxy environments, their halo GC populations show
some striking similarities. Perhaps the properties of halo GC systems
are determined by processes which are largely decoupled from those
which shape the main components of galaxies as we see them today.

 \section{Summary} 

 We have investigated the global properties of the M31 GC system using
 an updated sample which includes newly-discovered GCs reported by us
 in Paper I as well as other revisions to the RBC. We also derive
 structural parameters for 13 ECs. We find that many of these are less
 luminous and less extended than those presented in
 \citet{Huxoretal05}, bridging a gap between them and the
 ``Palomar-type" GCs found in the MW.

 \citet{Mackeyetal10b} recently showed that many clusters in the outer
 regions of M31 are physically associated with tidal streams, and the
 results presented in this paper are entirely consistent with this
 scenario.  Specifically, we find no evidence for a significant radial
 colour/metallicity gradient at large galactocentric radii, as
 expected from accretion. We also find evidence for a flattening in
 the GC number density radial profile in M31 occurring at a projected
 radius of $\sim$ 30 kpc, coinciding with a similar feature in the
 underlying stellar halo component. \citet{Abadietal06} have shown
 that such a flattening occurs naturally in galaxies that grow through
 a combination of in situ star formation and accretion, with the point
 of transition indicating the radius beyond which the bulk of the
 matter has been accreted.

 Wherever possible, we have compared the properties of the M31 halo GC
 system to that of the MW, often finding marked differences.  Although
 the overall form of the luminosity functions is similar in both
 systems down to M$_{V0}\approx -5$, M31 possesses a significant
 population of luminous and compact GCs at large galactocentric radii
 which, aside from NGC2419, have no counterpart in the MW. M31 also
 has a number of extended GCs, many of which are far larger than those
 in the MW (Figure \ref{Fi:mv_rh}).  On the other hand, halo GCs in
 M31 and the MW have similarily blue mean colours beyond
 R$_{proj}>15-30$ kpc, with little dispersion, indicating that old
 metal-poor populations dominate in both cases.  We suggest that the
 differences between the two GC systems could be, at least partly,
 explained by the differing accretion histories that M31 and MW have
 experienced.

 Finally, our work illustrates the importance of extending study of GC
 systems to large galactocentric radii. The accretion of
 cluster-bearing satellites is likely the dominant process in building
 up the halo GC populations of galaxies, while in situ formation may
 contribute the most to the populations at smaller radii.  As a
 result, it is the halo GC populations that have the most to tell us
 about the hierarchical assembly of galaxies. Large area surveys with
 telescopes such as Pan-STARRS and LSST should contribute
 significantly to building samples of GC candidates in the remote
 outskirts of galaxies within and beyond the Local Group, while
 spectroscopic campaigns will become increasingly necessary to weed
 out contaminants in these sparsely-populated parts.

\section*{Acknowledgments}

APH and AMNF were supported by a Marie Curie Excellence Grant from the
European Commission under contract
MCEXT-CT-2005-025869. ADM is also grateful for support by an Australian 
Research Fellowship from the Australian Research Council.
NRT acknowledges a STFC Senior Research Fellowship.  The Isaac Newton
Telescope is operated on the island of La Palma by the Isaac Newton
Group in the Spanish Observatorio del Roque de los Muchachos of the
Instituto de Astrof\'{i}sica de Canarias. This research also used the
facilities of the Canadian Astronomy Data Centre operated by the
National Research Council of Canada with the support of the Canadian
Space Agency.  Based on observations obtained with MegaPrime/MegaCam,
a joint project of CFHT and CEA/DAPNIA, at the Canada-France-Hawaii
Telescope (CFHT) which is operated by the National Research Council
(NRC) of Canada, the Institute National des Sciences de l'Univers of
the Centre National de la Recherche Scientifique of France, and the
University of Hawaii.  We also thank the anonymous referee, whose
comments greatly improved the quality of this paper.

\end{document}